\documentstyle[11pt,newpasp,twoside,epsf]{article}
\def\kpc{\rm kpc}
\def\au{\rm AU}

\def\rjup{R_{Jup}}

\def\edcomment#1{\iffalse\marginpar{\raggedright\sl#1\/}\else\relax\fi}
\reversemarginpar

\begin{document}
\title{Survey for Transiting Extrasolar Planets in Stellar Systems (STEPSS): The Frequency of Planets in NGC 1245}
\author{Christopher J. Burke$^1$, D. L. DePoy$^1$, B. Scott Gaudi$^2$, J. L. Marshall$^1$, Richard W. Pogge$^1$}
\affil{$^1$Department of Astronomy, The Ohio State University, Columbus, OH 43210-1173}
\affil{$^2$Hubble Fellow, School of Natural Sciences, Institute for Advanced Study, Einstein Drive, Princeton, NJ, 08540}

\begin{abstract}
We present first results from the Survey for Transiting Extrasolar
Planets in Stellar Systems (STEPSS). Our goal is to assess the
frequency of close-in extrasolar planets around main-sequence stars in
several open clusters.  By concentrating on main-sequence stars in
clusters of known (and varied) age, metallicity, and stellar
density, we will gain insight into how these various
properties affect planet formation, migration, and survival.  We show
preliminary results from our 19 night photometric campaign of the old,
solar metallicity cluster NGC 1245.  Taking into account the
photometric precision, observational window function, transit
probability, and total number of stars monitored, we estimate that
we should be able to probe planetary companion fractions of $<1\%$ for
separations of $a<0.03$ AU.  If $1\%$ of the stars in the cluster have
Jupiter-sized companions evenly distributed in $\log{a}$ between 0.03
and 0.3 AU, we expect to find $\sim 2$ transits.  A preliminary search
of our light curve data has revealed a transit with a depth $\sim 4\%$.
Based on its shape, it is likely to be a grazing binary eclipse
rather than a planetary transit, emphasizing the need for
high temporal resolution in transit surveys.
\end{abstract}

\section{Introduction}

The recent detections of candidate low-mass companions to local disk
stars via transits by several groups has verified the possibility of
discovering statistically significant numbers of extrasolar planets
through ground based transit surveys (Udalski et~al.\ 2002,
Mallen-Ornelas et~al.\ 2003).  Although such surveys will provide
valuable information about the types of planetary systems that exist,
it will be difficult to piece together the dynamical, chemical, and
evolutionary history of the parent stars.  The parent stars' physical
properties are particularly important since radial velocity surveys for
low-mass companions have revealed the trend of the fraction of stars
with planets increases as a function of the metallicity of the parent
star.  The interpretation of this trend is unclear and more
information is needed.

Simple stellar systems, such as globular clusters and open clusters,
are an excellent laboratory for transit surveys.  Such fields provide
$\sim 10^{3-5}$ stars of the same age and metallicity.  Interpretation
of the null result for the metal-poor globular cluster
47 Tuc (Gilliland et~al.\ 2000) has been complicated by the possible
effects of the dense stellar environment on planet formation and
survival.  Therefore, we are concentrating on metal-rich, sparser open
clusters. By observing main-sequence stars in 4-5 open clusters of
known (and varied) age, metallicity, and stellar density, we will gain
insight into how these various properties affect planet formation,
migration, and survival. Our primary instrumentation is the MDM
8192x8192 4x2 Mosaic CCD imager.  In combination with the MDM 2.4m
telescope it yields a 25x25 arcmin$^{2}$ field of view with
0.18"/pixel.  The Ohio State University has access to a
25\% share of the MDM facility; in 2001 we were granted 19 nights on
the 2.4m, we have been granted $\sim 60$ nights during Fall 2002,
and we expect to
obtain 20-40 nights for each cluster in the future with the same
instrumental setup.

\begin{figure}
\plottwo{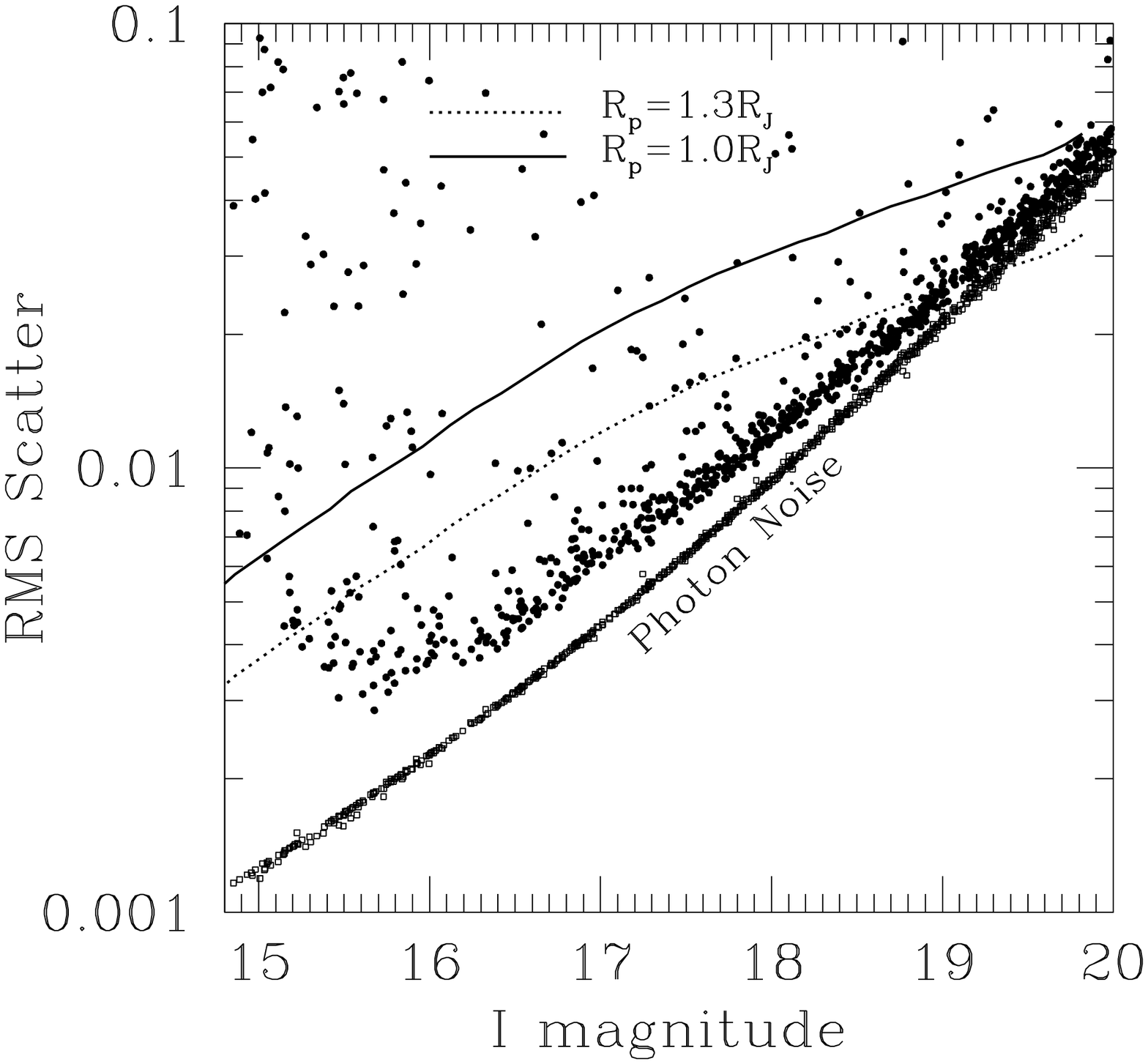}{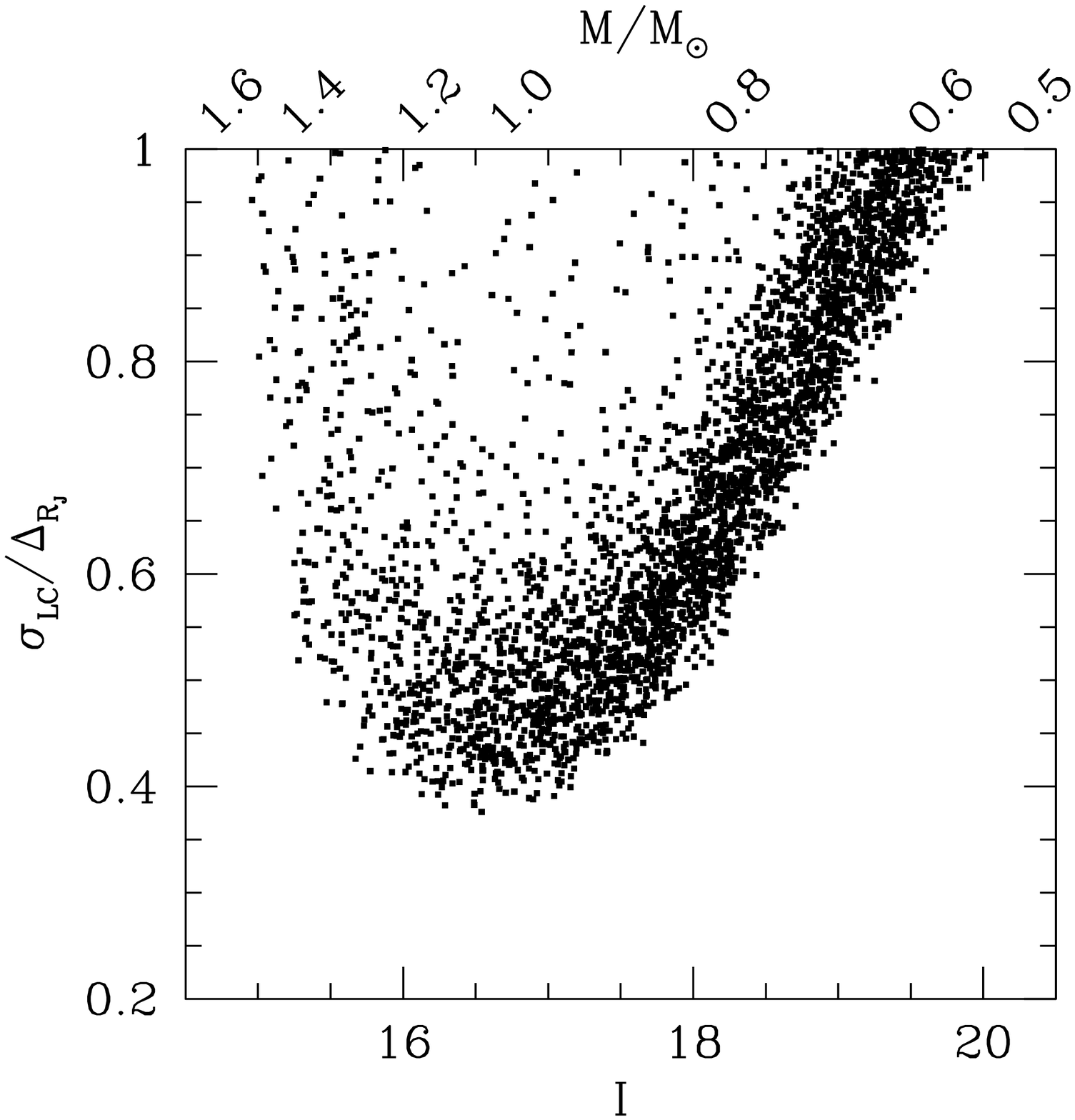}
\caption{(a) RMS scatter over the entire run as a function of $I$
magnitude for all stars on a single CCD chip.  (b) The ratio of
the transit depth due to a 1 $\rjup$ planet to the RMS scatter.
}
\end{figure}

\begin{figure}
\plotfiddle{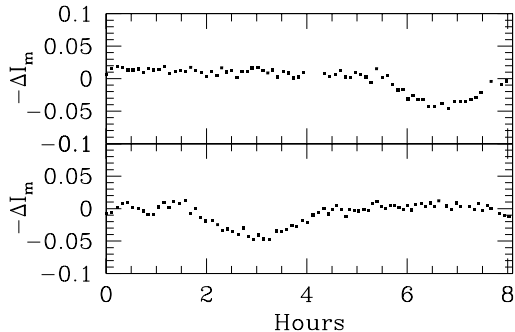}{4.0cm}{0}{35}{35}{-120}{-90}
\caption{light curves for a transit candidate observed on two separate nights.
Unfortunately, this is likely to be a grazing binary eclipse. }
\end{figure}

\section{Observations, Analysis and Preliminary Results for NGC 1245 }

In Oct. 2001 we obtained data over 19 nights for the open cluster NGC
1245, an old ($\sim 1$ Gyr), solar-metallicity cluster with $\sim 1500$
members.  We lost 4 nights to inclement weather and the remainder were
non-photometric, resulting in a total of 960 $I$-band images. 
We performed the photometric reduction using the DoPhot PSF fitting 
package.  Differential light
curves for individual target stars were derived by 
locating 13 optimal comparison stars close to each target.  We plan to 
test difference imaging techniques in the future.  The points
in Figure 1(a) show the RMS scatter in the light curves as a
function of I-magnitude.  The lower, tight sequence of points in Figure 1(a)
shows the expected photon noise.  At the faint end, the RMS scatter is only
$\sim 15\%$ larger than the photon limit.  At the bright end, there
is a systematic error $\sim 0.26\%$, which we suspect is due to
flat-fielding errors.  The lines in Figure 1(a) represent the transit
depth for a 1.3 and $1.0\, \rjup$ planet.  Figure
1(b) shows the ratio between the transit depth of a $1.0\, \rjup$ planet and
the light curve RMS scatter.  There are $\sim 4300$ stars with
sufficient $S/N$ to detect a $\rjup$ transit.  Also shown in this figure is
the estimated mass of the parent star assuming a 1 Gyr, solar-metallicity
population at 2.5 $\kpc$ and E(B-V)=0.26.
We estimate $35\%$ of these stars to be cluster members.  

A preliminary search through the
data has revealed a transit candidate with a depth $\sim 4\%$.
We observed the transit on two
separate nights and the light curves 
are shown in Figure 3.  Unfortunately, the transit is unlikely
to be a planetary transit, since the transit shape is
indicative of a grazing binary eclipse.  Planetary transits tend to
be boxier in shape with more rapid ingress and egress durations and
flat-bottom eclipses (Seager \& Mallen-Ornelas 2003).  
However, this example demonstrates the
extreme importance of sufficient time resolution in transit surveys
in order to easily discard signals that can be
confused with planetary transits.

\begin{figure}
\plottwo{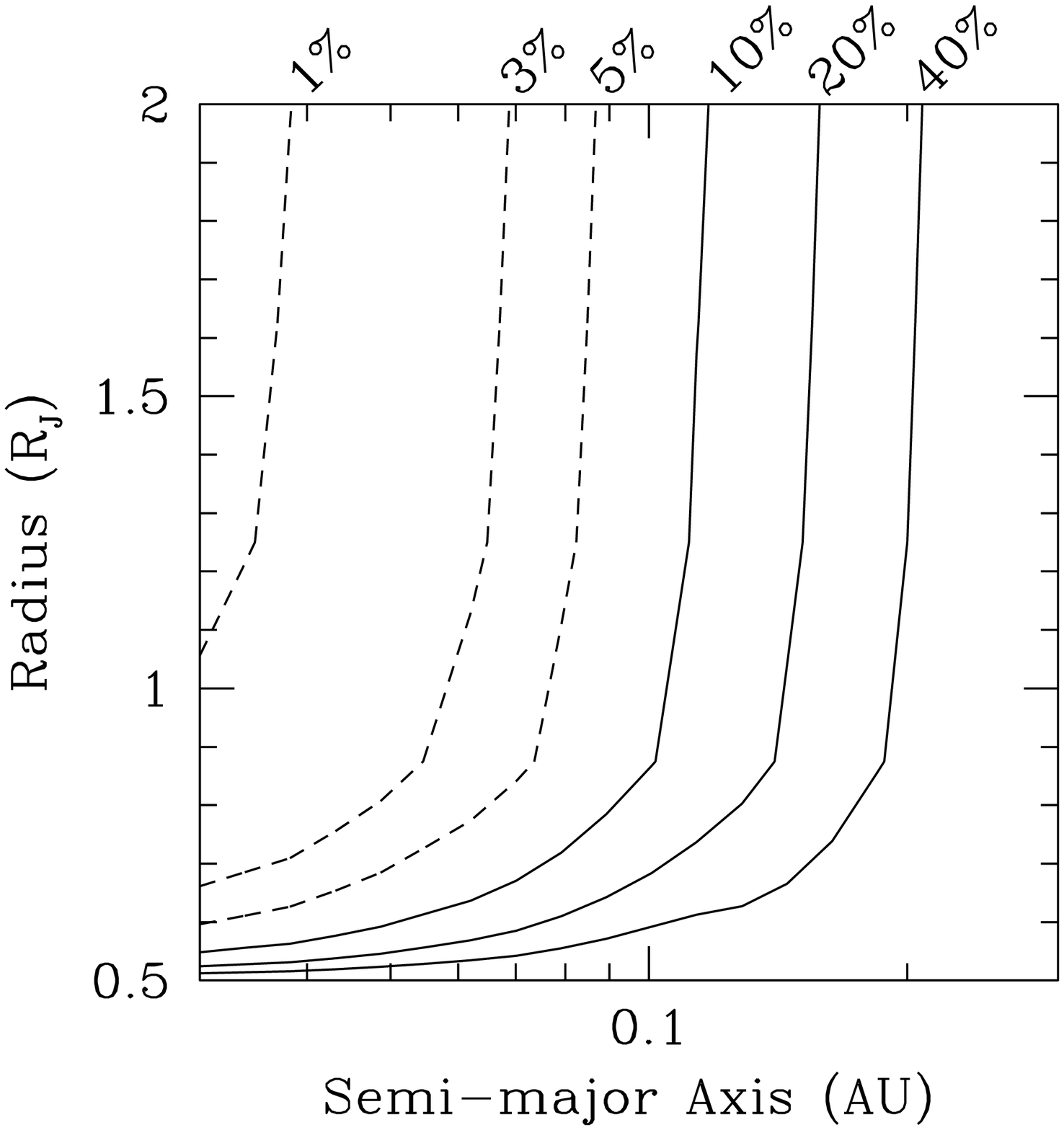}{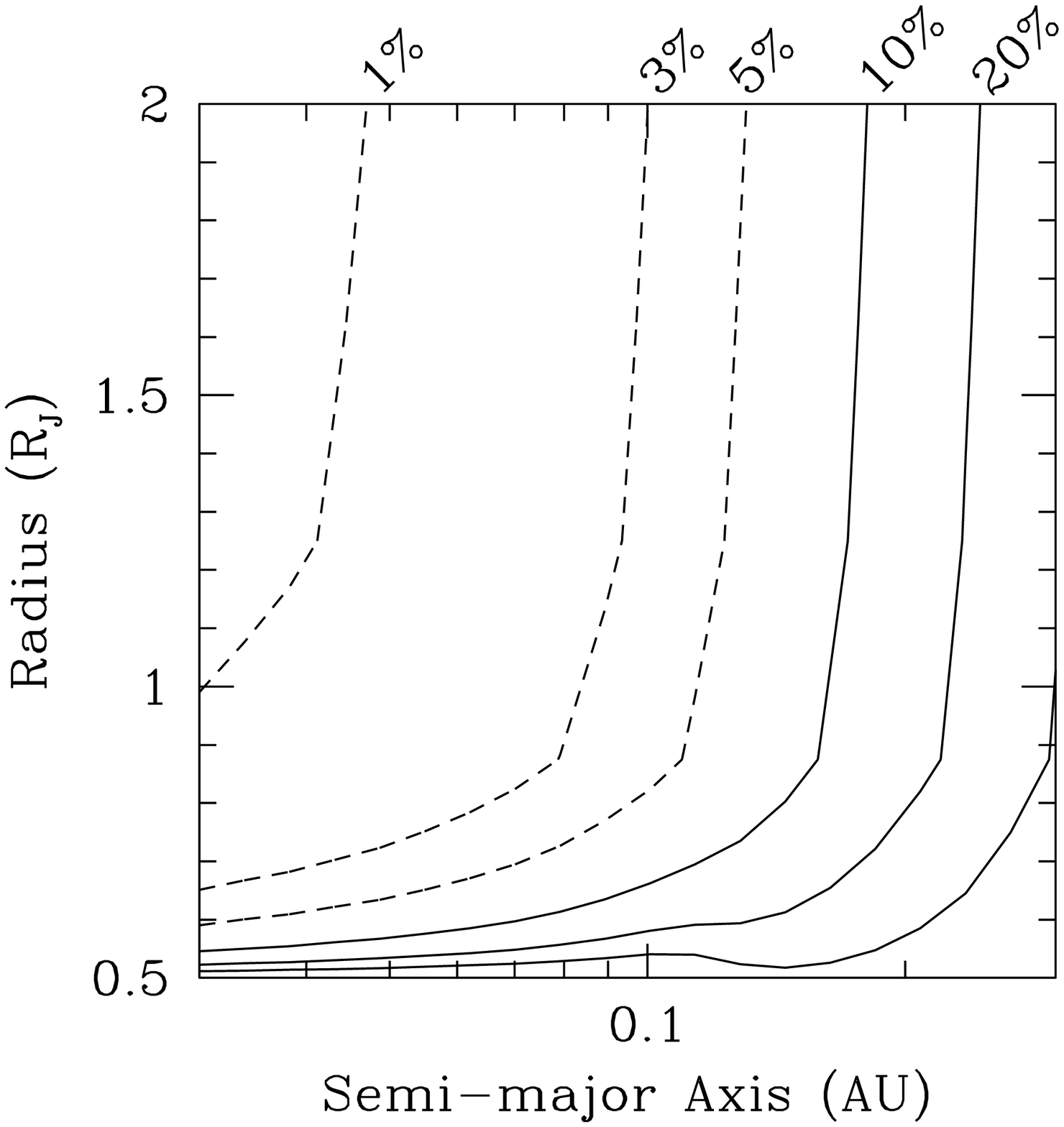}
\caption{(a) Expected $95\%$ upper limits on
fraction of stars with planets for NGC 1245. (b) Theoretical upper limits
if we had observed NGC 1245 for 40 nights, with 8 hours each night,
using the identical instrumental setup.
}
\end{figure}

\section{Expected Transit Results and Future Work}

We calculate our sensitivity to planetary transits using the formalism 
outlined in Gaudi (2000).  Taking into account the
cluster luminosity function, distance, and reddening, our observational
photometric sensitivity and window function, and assuming that $1\%$
of stars have $\rjup$ companions evenly distributed in log
semi-major axis between 0.03 and 0.3 AU, we expect to detect $\sim 2$
transits in the NGC 1245 data set.  The curves in Figure
3(a) show the $95\%$ upper confidence limits we can place on the
fraction of stars with planets for a given configuration of planetary
companion separation versus radius.  For instance, based on our data
obtained for NGC 1245 and a null result, we can exclude greater
than $10\%$ of the cluster members having a Jupiter-sized object
orbiting within $0.1\au$.  Figure 3(b) shows the theoretical limits
if we had observed NGC 1245 for 40 nights, with 8 hours each night,
using the identical instrumental setup.  Such results may be achieved
with our upcoming 60 nights (30 on MDM 2.4m and 30 on MDM 1.3m) to
observe NGC 2099 which is 2.5 times closer than, and has comparable number
of members to NGC 1245.

We have demonstrated that we can easily achieve the
photometric precision necessary to detect Jupiter-sized transits
around main-sequence stars.  In light of the uncertain radial velocity
results of the fraction of stars with planets as a function
of metallicity, our theoretical expectations show that even without
the discovery of planetary transits our survey will place interesting
constraints on this relation for a population of determinable
metallicity, age, and stellar density.  
We emphasize the importance of sufficient time
resolution in transit surveys, as has been demonstrated by other surveys 
(Mallen-Ornelas et~al.\ 2003).  We will continue to optimize our transit detection
methods to search for lower amplitude ($\la 1\%$) transit signals.

\acknowledgements

This work was supported by NASA through a Hubble Fellowship grant
from the Space Telescope Science Institute, which is operated by the
Association of Universities for Research in Astronomy, Inc., under
NASA contract NAS5-26555.

\end{document}